# Experimental investigation of a coherent quantum measurement of the degree of polarization of a single mode light beam

M. LEGRE, M. WEGMULLER, N. GISIN
Group of Applied Physics, University of Geneva, 1211Geneva 4, Switzerland

***Abstract**: A novel method for the direct measurement of the degree of polarization is described. It is one of the first practical implementations of a coherent quantum measurement, the projection on the singlet state. Our first results demonstrate the successful operation of the method. However, due to the nonlinear crystals used presently, its application is limited to spectral widths larger than ~8nm.*

## I. Introduction

The degree of polarization (DOP) of a light beam is a parameter of fundamental importance in many applications. Typically, either a DOP close to 1 (totally polarized light, e.g. for interferometric measures), or a DOP as small as possible (depolarized, no polarization dependence) is desired. Intermediate values of the DOP (or partially polarized light) on the other hand are frequently generated through depolarization in optical systems, and its measurement can consequently be exploited for system characterization. In a telecom link e.g., the light can become depolarized due to the dispersion among the two polarization modes (PMD). This detrimental signal degradation can be minimized by inserting a PMD compensator, where the DOP is routinely used as a feed-back parameter to control the dynamic adjustment of the compensator [1,2]. For such applications, it is interesting to have a compact, fast, and cheap measurement apparatus to determine the DOP. An interesting approach was recently suggested by one of the authors [3]. He demonstrated that the DOP is directly linked to the projection onto the so-called singlet state ($y^{(-)} = \left(|H_1,V_2\rangle - |V_1,H_2\rangle\right)/\sqrt{2}$), one of 4 specific 2-photon polarization states (Bell states). $H_1$ corresponds to the first photon being horizontally polarized, $V_2$ to the second photon with vertical polarization, etc. Bell state measurements are in fact an important tool in quantum physics, and are e.g. employed to characterize teleportation experiments [4]. Note that it is also possible, as in a classical polarimeter, to determine the polarization state from the projection on the four Bell states.

The principle of the DOP measurement by projection on the singlet state is relatively simple. In fact a singlet state projection can be realized by using nonlinear interactions, as demonstrated by *Kim et al* [5]. However, with respect to a potential application as a DOP meter for telecoms, some special points need to be addressed. First, there exist three different causes of depolarization of a light beam: time mixture (the polarization fluctuates with time), spatial modes mixture, and wavelengths mixture. In telecoms, the beam is typically monomode, and the nonlinear interactions used in our DOP meter are quasi-instantaneous, so what will actually be measured is the depolarization due to the spectral bandwidth of the source (wavelengths mixture) alone. This distinguishes our DOP meter from a classical polarimetric DOP measurement, where both wavelength and time mixture (due to the finite detector integration time) are monitored. Second, working at a wavelength of 1.5 um is not very common for quantum measurements as a proper detection is much harder to achieve. Third, due to their bosonic nature, the projection on the singlet state of two photons in the same mode (spatial and temporal) having the same wavelengths becomes null. In [5], this caused no problem as a large wavelength separation of 100nm was used. For telecom applications on the other hand, the channel bandwidth is only around 0.3nm (40Gb/s system). The question of interest, investigated in our paper, is therefore whether the singlet projection will still work for such small wavelength separations.

The paper is organized as follows. In section I, the principle of the singlet state measurement to be employed is described in detail. Especially, the different cases of very large, zero, and small wavelength separations are treated, allowing to understand that the quality of the singlet projection will gradually decrease with smaller separations. After this theoretical consideration, we characterize the employed nonlinear crystals (section III), and verify our model by performing singlet projection measurements for three different wavelength separations (part IV).

## II. Principle of the projection on the singlet state

Before analyzing the influence of the wavelength separation on the singlet projection quality, we briefly recapitulate the principle of operation as described in [3].

As depicted in figure 1, the projection on the singlet state is realized as follows. The weight of $\left|H_1, V_2\right\rangle$ is determined by the sum frequency generation (SFG) in a type II crystal, where the phase matching favors the two-photon process with the lower frequency photon polarized vertically and the higher frequency one horizontally. Accordingly, the weight of $\left|V_1, H_2\right\rangle$ is determined in the second crystal, identical to the first one but rotated by 90° (see figure 1). Finally, the projection on $\psi^{(-)}$ is realized by making interfere -with a phase of $\pi$- the two orthogonal SFG photons with a linear polarizer at 45°.

In order to see why and when the wavelength separation becomes limiting, we now give a more detailed description than was necessary in [3]. We will consider two photons with wavelengths $\lambda_1$ and $\lambda_2$, respectively. Each photon has its own polarization and a DOP of 1 (completely polarized). For simplicity of notation, we allow for linear polarizations only, and define the two polarizations with two angles $\theta$ and $\varphi$ in the plane (x,y), with x parallel to the X axis of the first crystal (fig.2). $\theta$ is the angle between the x axis and the polarization of $\lambda_1$, and $\varphi$ the angular difference between the two polarizations (i.e. $\lambda_2$ is polarized along $\theta+\varphi$). The DOP of such a source is given by $\sqrt{(I_1 + I_2)^2 - 4I_1 I_2 \sin^2\varphi} \big/ (I_1 + I_2)$, and consequently essentially depends on the relative polarization states of $\lambda_1$ and $\lambda_2$.

A) Strongly separated wavelengths

We suppose there is phase-matching (PM) for the sum frequency generation (SFG) in crystal 1 only for $\lambda_1$ parallel to x and $\lambda_2$ parallel to y ($\theta$ = 0°, $\varphi$ = 90°), and that the polarization of the created wave is parallel to y. The second crystal is identical to the first, but rotated by 90°. Consequently, the waves generated in crystals 1 and 2 are given as

$$\vec{E}^{(1)}_{\omega_1+\omega_2} = E_1 E_2 \cos\theta \sin(\theta + \varphi)\vec{y} \qquad (1)$$

$$\vec{E}^{(2)}_{w_1+w_2} = -E_1 E_2 \sin q \cos(q + j)\vec{x} \qquad (2)$$

and the overall intensity after projection on the 45° direction (linear polarizer) becomes

$$I = \frac{(E_1 E_2)^2}{2}\sin^2 j \qquad (3)$$

The intensity is θ independent, reflecting the rotational invariance of the singlet state. Note that the same result is found for elliptical polarizations, only that φ becomes the difference between $\chi_1$ and $\chi_2$ ($\tan\chi = E_y/E_x$). Consequently, if we project two photons on the singlet state, only their relative, but not their absolute, polarizations are important. This is what is exploited for our DOP-meter. Comparing the measured intensity (eq.3) to the DOP of our source (see above), one finds that *I* is a direct measure of ($1-DOP^2$) (as has been demonstrated in [3] for any kind of sources).

B) Equal wavelengths

What happens when the two wavelengths are equal? Obviously, the previous assumption that there is only PM if $\lambda_1$ parallel to x and $\lambda_2$ parallel to y no longer holds for a case where $\lambda_1=\lambda_2$ - the above PM condition inherently leads to PM for $\lambda_1$ parallel y and $\lambda_2$ parallel x as well. In fact, SFG can no longer be distinguished from the second-harmonic generation SHG. Analog to the previous section, the waves generated in crystals 1 and 2 are

$$\vec{E}^{(1)}_{w_1+w_2} = \left[\frac{E_1^2}{2}\sin 2q + \frac{E_2^2}{2}\sin(2(q + j)) + E_1 E_2 \sin(2q + j)\right]\vec{y} \qquad (4)$$

$$\vec{E}^{(2)}_{w_1+w_2} = -\left[\frac{E_1^2}{2}\sin 2q + \frac{E_2^2}{2}\sin(2(q + j)) + E_1 E_2 \sin(2q + j)\right]\vec{x} \qquad (5)$$

If we project the total amplitude on the 45° direction again, the result is zero (I=0), and consequently DOP=1. This result is not surprising, as it is well known that strictly monochromatic light is completely polarized. The result can also be understood from the bosonic nature of the photons - a projection of their symmetric spatial and frequency state on the anti-symmetric singlet is null.

C) Similar wavelengths

The question of interest is now if the DOP can also be extracted from the singlet-projection for the telecom relevant case of similar but non-equal wavelengths.

As was indicated in the previous sections, for a ‘good’ projection on the singlet state (using collinear PM), the PM should allow only for one kind of SFG (e.g. $\lambda_1$ ordinary (i.e. parallel x) and $\lambda_2$ extraordinary). While this can be realized well for strongly separated wavelengths, the SFG for the opposite case (PM for $\lambda_1$ extraordinary and $\lambda_2$ ordinary) can no longer be neglected for $\lambda_2 \to \lambda_1$. The influence of this second, perturbing process will become important when the difference of PM angles for these two SFG becomes smaller than the angular bandwidth of the crystal, $\delta\theta_{1/2}$ (twice the angular deviation from the PM angle for which the SFG efficiency is reduced by a factor of 2). For these cases, one must consider a superposition of the two situations presented above. If we weight the SFG efficiency for $\lambda_1$ extraordinary and $\lambda_2$ ordinary with 1 and the opposite, perturbing case with $\alpha<1$, we have from the first crystal

$$\vec{E}^{(1)}_{\omega_1+\omega_2} = \big[\underbrace{E_1 E_2 \cos\theta \sin(\theta+\varphi)}_{\text{process } \lambda_1 \text{ e \& } \lambda_2 \text{ o}} + \underbrace{\alpha E_1 E_2 \sin\theta \cos(\theta+\varphi)}_{\text{process } \lambda_1 \text{ o \& } \lambda_2 \text{ e}}\big]\vec{y} \qquad (6)$$

which can be written as

$$\vec{E}^{(1)}_{\omega_1+\omega_2} = E_1 E_2 \big[(1-\alpha)\cos\theta \sin(\theta+\varphi) + \alpha \sin(2\theta+\varphi)\big]\vec{y} \qquad (7)$$

In analogy, the SFG in the second crystal is

$$\vec{E}^{(2)}_{\omega_1+\omega_2} = E_1 E_2 \big[(1-\alpha)\sin\theta \cos(\theta+\varphi) + \alpha \sin(2\theta+\varphi)\big]\vec{x} \qquad (8)$$

The overall intensity after the 45° projection thus becomes

$$I = \frac{E_1^2 E_2^2}{2}(1-\alpha)^2 \sin^2\varphi \qquad (9)$$

This is the main result of this section. It demonstrates that the intensity is still proportional to $\sin^2\varphi$ (i.e. to 1-DOP$^2$), but with an amplitude reduced by a factor of $(1-\alpha)^2$.

We now see that for decreasing wavelength separations, where $\alpha$ approaches 1, the signal of interest becomes smaller and smaller. Evidently, below a certain wavelength difference, the projection becomes completely immersed in the measurement noise, so that one can no longer gain any information on the DOP. This minimal wavelength separation consequently depends on the crystals (SFG efficiency, angular bandwidth) and on all the different elements of the set-up leading to noise. Accordingly, in the next section, we will describe the crystals chosen for our experiment and determine the weighting factor $\alpha$.

## II. Choice and characterization of the nonlinear crystal

For a well collimated beam and a source with a narrow spectral bandwidth, we have seen in the previous section that the PM acceptance, i.e. the angular bandwidth $\delta\theta_{1/2}$, becomes important. Besides a small $\delta\theta_{1/2}$, the efficiency of the SFG process should be as large as possible for a good signal-to-noise ratio. Although they are not very selective (see below), we chose KTP type II crystals (XZ plane) as they promise for a good SFG efficiency and are readily available. We opted for a short crystal length of just 3mm, in order to have a good mode overlap (i.e. conversion efficiency) in *both* crystals - note that the spatial walk-off between the ordinary and extraordinary beam is quit important in KTP (3°). The angular bandwidth $\delta\theta_{1/2}$ for this crystal type and length was calculated [6] to 0.37°. The influence of the perturbing SFG should be negligible if the difference among the PM angles of desired and perturbing SFG (again calculated from [6]) become larger than $\delta\theta_{1/2}$. The situation is shown in Fig.3 for the case $\lambda_1$=1542 nm and $\lambda_2>\lambda_1$. According to the figure, a wavelength separation of 18 nm or more ($\lambda_2\geq$1560 nm) should allow for an unperturbed projection on the singlet state (it allows for a good suppression of the undesired SFG process). For a wavelength separation of 8 nm however, both SFG processes will co-exist, and the measurement risks to be perturbed by the noise. A lower bound for the wavelength separation can be defined by requesting a minimal SNR of 1, where the singlet intensity (eq.9) equals the noise. Consequently, both the weighting parameter $\alpha$ (eq.6) and the measurement noise have to be known. The latter depends on a variety of parameters and includes more than just the thermal noise of the detector. It requires knowledge of the experimental set-up, and will be further discussed at the end of the paper.

The weighting factor $\alpha$ is determined experimentally. To do so, we utilize one crystal only, and compare the power from SFG obtained for the desired ($\lambda_1$=1542 nm adjusted as ordinary, $\lambda_2$ as extraordianry wave) and the perturbing, reversed case. This ratio is equal to $\alpha^2$, and is shown in Fig. 4 for $\lambda_2$= 1542-1560nm. As expected, $\alpha^2$ tends to 1 when the difference decreases and becomes negligible above $\lambda_2$=1560nm. In fact, from Eq.9, $(1-a)^2$ can be considered as the relative efficiency of the singlet projection.

## III. Experimental realization of the singlet projection

The set-up used for the singlet projection is shown in Fig. 5. The source is composed of two tunable lasers around 1550nm, combined by a fiber coupler. Each polarization can be controlled independently. The total power after amplification is about 70mW. Before the bulk part of the set-up, the light is collimated with a fiber GRIN lens. The beam waist must be small for a good efficiency of the SFG, yet large enough for good collimation in both crystals. With respect to the first crystal, the second is rotated by 90° along the direction of beam propagation (Y axis). Two birefringent plates are then used to adjust the phase between the two SFG signals from the two crystals. This is done by tilting the plates (whose birefringence axes are aligned with the crystal axes) in opposite directions, thereby avoiding any spatial beam displacement that would necessitate a re-adjustment of the detection. The singlet state is then selected by a linear polarizer at 45° with respect to the crystal axes. A pinhole serves to increase the spatial coherence of the light. Finally, a monochromator is used to both select the photons created by SFG and to reject the perturbing signal from SHG. The so-filtered light is then detected with a silicium photodiode operated in photon counting mode. Further, to analyse the correct operation of the DOP meter, a polarization analyzer could be inserted in front of the crystals (dashed elements in Fig. 5).

As discussed previously, one laser ($\lambda_1$) was set to 1542 nm, whereas the other was adjusted for a wavelength separation of 18nm, 8nm, and 4nm, respectively ($\lambda_1<\lambda_2$). The results of the singlet-projection for these three cases are shown in Figs.7-9, where the number of counts are given as a function of the polarization difference $\varphi$ between the two lasers. The different squares for one specific angle $\varphi$ correspond to different absolute polarization orientations $\theta$, whereas the cross gives their mean values.

The (ideally absent) variation of the count-rate with $\theta$ is essentially caused by a misalignment of the birefringent plates, and by the non-complete (i.e. partial) spatial coherence. This was determined with a study on the influences of the different misalignments in our setup. In our model, we varied the rotation angle of the second crystal, the birefringent axis direction of the birefringent plates, and the alignment of the 45° analyzer. Finally, the visibility of the interference was allowed to be reduced as well (this can arise e.g. from an insufficient spatial filtering). To investigate the relative weight of these misalignments, the standard deviation of the measured singlet intensity for different θ is calculated as a function of φ. The corresponding results are shown in Fig. 6, where for each plot, one of the above parameters was moved from the optimum position by either 1° for the angular alignments, or reduction in visibitlity to 0.9. Note that the chosen values correspond to what we can expect for our measurment set-up. The figure demonstrates that the visibility is the most crucial alignment parameter for φ close to zero (the point of interest in telecom systems). For φ tending to 90°, it is the misalignment of the birefringent plates that becomes most important. The influence of the two others parameters can not be completely neglected, but are of less importance. Consequently, to improve the quality of the singlet-projection, we need to be especially careful with the visibility and the alignment of the birefringent plates.

Coming back to the experimental results of the singlet-projection, and concentrating on the first measurement with a wavelength difference of 18 nm (Fig.7), one sees that the closer φ gets to 0, the smaller the number of counts. This is the expected behavior where the number of counts is proportional to (1-DOP$^2$). However, the θ dependence of the result reduces the accuracy of a single measurement, and one should therefore rather use the mean value from several measurements with different θs. As the figure demonstrates, this mean values follow the predicted theoretical behavior (fit with the function $a \cdot \sin^2(\varphi) + b$, Eq.9) very well.

As a measure for the quality of the DOP measurement, one can use the visibility $V = \frac{N_{\max} - N_{\min}}{N_{\max} + N_{\min}}$, with $N_{max}$ and $N_{min}$ the maximum and minimum mean count rate, respectively. For the present measurement, V=93%, indicating that the DOP measurement works well for a wavelength separation of 18nm.

In Fig.8, the results for a wavelength separation of 8 nm ($\lambda_1$=1542nm, $\lambda_2$=1550nm) are presented. Note that for simplicity, only the points for positive φ have been measured. The results are qualitatively the same as for 18 nm wavelength separation, but the visibility V is reduced to 85%, and the fluctuations for different θ are slightly larger. In fact, we have seen in the theoretical part that the efficiency of the projection on the singlet state decreases with $(1-\alpha)^2$, making measurements with smaller wavelength separations (i.e. a larger relative amplitude α of the perturbing SFG process) more sensitive to the noise. Although a wavelength separation of 8 nm is smaller than the limit for completely unperturbed operation of the singlet projection for the present crystals, the good visibility clearly indicates that a reasonable DOP extraction can still be achieved.

The results become qualitatively different for a wavelength separation as small as 4 nm (Fig.9). The noise ‘fluctuations’ are very large now, covering almost completely the dependence on $\varphi$, i.e. the DOP. Nevertheless, the experimental data are still found to be very reproducible. The mean values are not constant as one might expect at first view, but are found to vary as $a + b\sin^2(\varphi) + c\sin^2(\varphi + 2\delta)$. The reason for this is revealed by a closer inspection of how Eq.9 was obtained. For it’s derivation, we assumed that the two terms $\alpha\sin(2\theta + \varphi)$ in (7) and (8) exactly compensate each other. This is however not completely true, as is indicated by a reduced visibility. For an $\alpha$ as large as in the present case, this flaw, leading to an additional noise term of $\alpha^2\sin^2(2\theta + \varphi)$, can no longer be neglected. Although this term would still vanish for averaging over all possible $\theta$, in the present case where only five different settings taken uniformly between 0 and 90° are used, a $(a + c.\sin^2(\varphi + 2\delta))$ dependence of the mean values persists.

**Discussion of the results**

We can define three domains of operation for the singlet projection. 1) $d\lambda > \Lambda_1$, the projection can be done without any problems. 2) $\Lambda_2 < d\lambda < \Lambda_1$, the measurement is increasingly noisy, but the DOP can still be extracted. 3) $d\lambda < \Lambda_2$, the signal is completely immersed in the noise.

The limit $\Lambda_1$ can be obtained from the angular bandwidth and the wavelength dependence of the phase-matching of the employed crystals, as discussed in section II. In the present case of 3mm long KTP type II crystals and $\lambda_1$=1542nm, $\Lambda_1$ =18 nm. The experimentally obtained visibility for this wavelength separation was found to be 93%. It dropped to ~80% for $d\lambda$=8 nm, which still allows for a proper DOP extraction from the singlet projection. Consequently, $\Lambda_2$ has to be smaller than 8 nm. The theoretical determination of $\Lambda_2$ is however difficult. This is because it depends on many different set-up parameters. If we consider a signal to noise ratio (signal: $\sin^2(\varphi)$, noise: $\sin^2(2\theta + \varphi)$) of 1 as the limiting criteria, a value of $\Lambda_2$ ~ 5.5nm is predicted for a visibility of 90%, in fairly good agreement with the experimental results of above (good operation for d$\lambda$=8 nm, no-go for d$\lambda$=4 nm).

Obviously, a minimum wavelength separation of 8 nm is largely insufficient for the targeted telecom applications. As shown above, reducing this value is merely a question of choosing the proper nonlinear crystal. However, this is not an easy task as the desired specifications are diametrically opposite to what is usually desired.

GaSe, HgS (Cinnabar), Banana, or POM are potential candidates as they have small (<1nm cm) spectral bandwidths. Unfortunately, these crystals are either hard to fabricate or hard to manipulate. Another way to reduce the spectral bandwidth would be to use longer lengths of standard crystals. E.g., 5 cm length KTP crystals would have the required specs, but, because of the spatial walk-off [6], the effective interaction length is just a few millimeters. It therefore requires walk-off compensation, which complicates the set-up.

**Conclusion**

The possibility of a DOP meter based on a coherent quantum measurement, the projection on the singlet-state, was experimentally investigated. Such a DOP meter is ultrafast, and would allow for both a very compact and low cost design. We tested our lab-apparatus using a light source of two discret wavelengths around 1.55 um, combined with adjustable relative polarization states to give any desired DOP.

For wavelength separations larger or equal to the angular bandwidth of the employed KTP crystals (18 nm), the DOP meter was found to perform well, with a reasonable precision (~5% for a DOP~1). The variation in the results is mainly due to a dependence on the absolute orientation of the polarization states, and could be further improved upon by, e.g., better spatial filtering. For smaller wavelength separations, the quality of the measurement gradually decreased, until it was no longer possible to extract any information on the DOP. This behavior was found to be caused by a perturbing SFG process. Besides the desired phase-matching condition (e.g. $\lambda_1$ ordinary, $\lambda_2$ extraordinary), the opposite one ($\lambda_1$ extraordinary, $\lambda_2$ ordinary) starts to be increasingly efficient. The above mentioned limit of operation can consequently be estimated by evaluating the wavelength separation at which the two processes become equally important. A value of ~5nm was found for the currently employed crystals, in good agreement with the experimental findings.

For the DOP measurement in telecom applications (e.g. feedback parameter for PMD compensator), this is clearly not sufficient as typically encountered wavelength separations are below ~0.3nm. However, the present minimum wavelength separation of our DOP meter is not a fundamental limit, but depends merely on the angular bandwidth of the employed nonlinear crystal. Although their high spatial walk-off would require some compensation scheme [7], cm long crystals of GaSe or HgS offer very narrow PM bandwidths in the order of 0.25nm. Also, waveguides could offer similar performances [8-9]. The feasibility of a DOP meter operating for wavelength separations below 0.3nm is therefore a matter of properly cutting and assembling the corresponding nonlinear crystals, which, however, is not mastered at this time.

***Acknowledgements**: Financial support from the Swiss OFES in the frame of the COST 265 project, EXFO Inc (Vanier, Canada), and the Swiss NCCR "Quantum photonics" are acknowledged.*

FIGURES

FIG. 1. Schematic of the principle of the singlet state projection. ↕ and ↔: vertical and horizontal orientation of the optical axis of the two crystals.

FIG. 2. Definition of the angles $\theta$ and $\varphi$. The directions x and y are parallel to the optical axis of the first and second crystal, respectively.

FIG. 3. Phase-matching angle versus $\lambda$ for the two possible SFG processes. The angular acceptance of the 3mm KTP crystals is 0.37°. Therefore, it is possible to select only one SFG process for wavelengths of 1542 and 1560nm, whereas for 1542 and 1550nm both SFG will co-exist.

FIG. 4. Ratio $\alpha^2$ of the power of the two SFG signals ( and ), as a function of the wavelength $\lambda_2$. As expected, the ratio tends to 1 for small wavelength differences. experimental data: squares, theoretical fit: solid line.

FIG. 5. Diagram of the experimental setup.

FIG. 6. Standard deviation of the singlet-state intensities for different $\theta$, as a function of the angle $\varphi$. Each curve is associated with a misalignment of 1° of one element of the setup, or a reduction of the visibility of 10%. Solid line: misalignment of the second crystal, dotted line: of the half-wave plate, dash-dotted line: of the polarizer, dashed line: reduction of the visibility.

FIG. 7. Measured intensity of the singlet-state as a function of the relative polarization angle φ for $\lambda_1$=1542nm and $\lambda_2$=1560nm. Square data points: results for different θs, crosses: mean values, solid line: fit with the function $a\sin^2(j)+b$.

FIG. 8. Measured intensity of the singlet-state as a function of the relative polarization angle φ for $\lambda_1$=1542nm and $\lambda_2$=1550nm. Symbols same as in figure 6.

FIG. 9. Measured intensity of the singlet-state as a function of the relative polarization angle φ for $\lambda_1$=1542nm and $\lambda_2$=1546nm. Square and bar data points: two different sets of measurements, solid line: fit with the function $a+b\sin^2(j)+c\sin^2(j+2d)$.

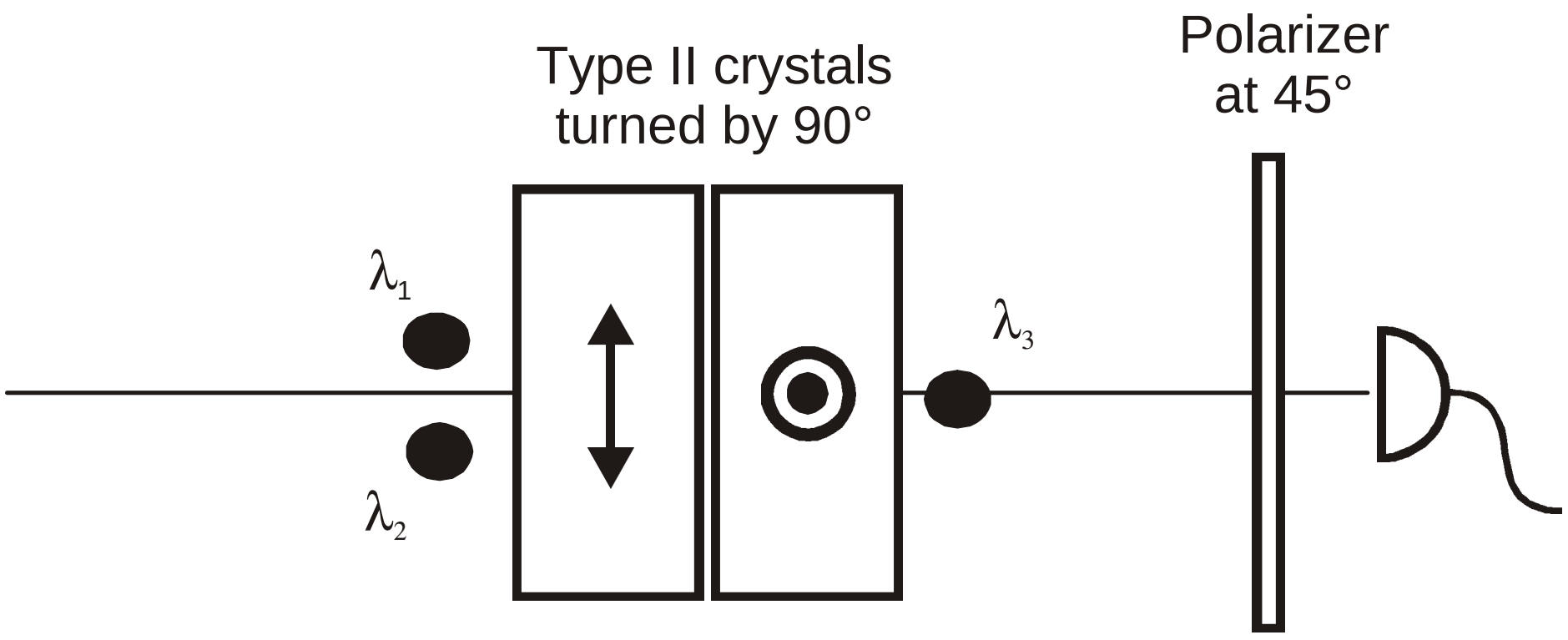

Figure 1

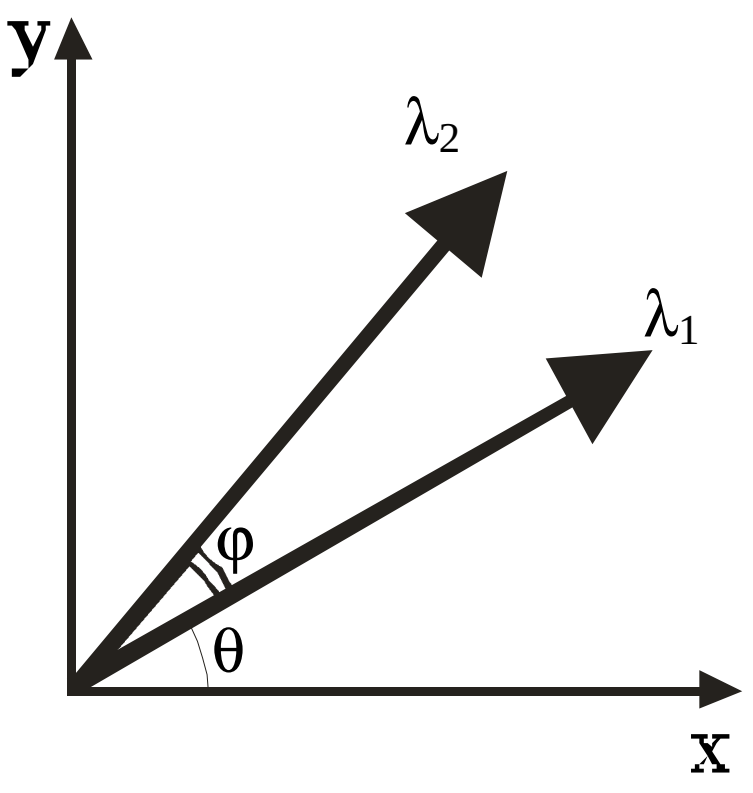

Figure 2

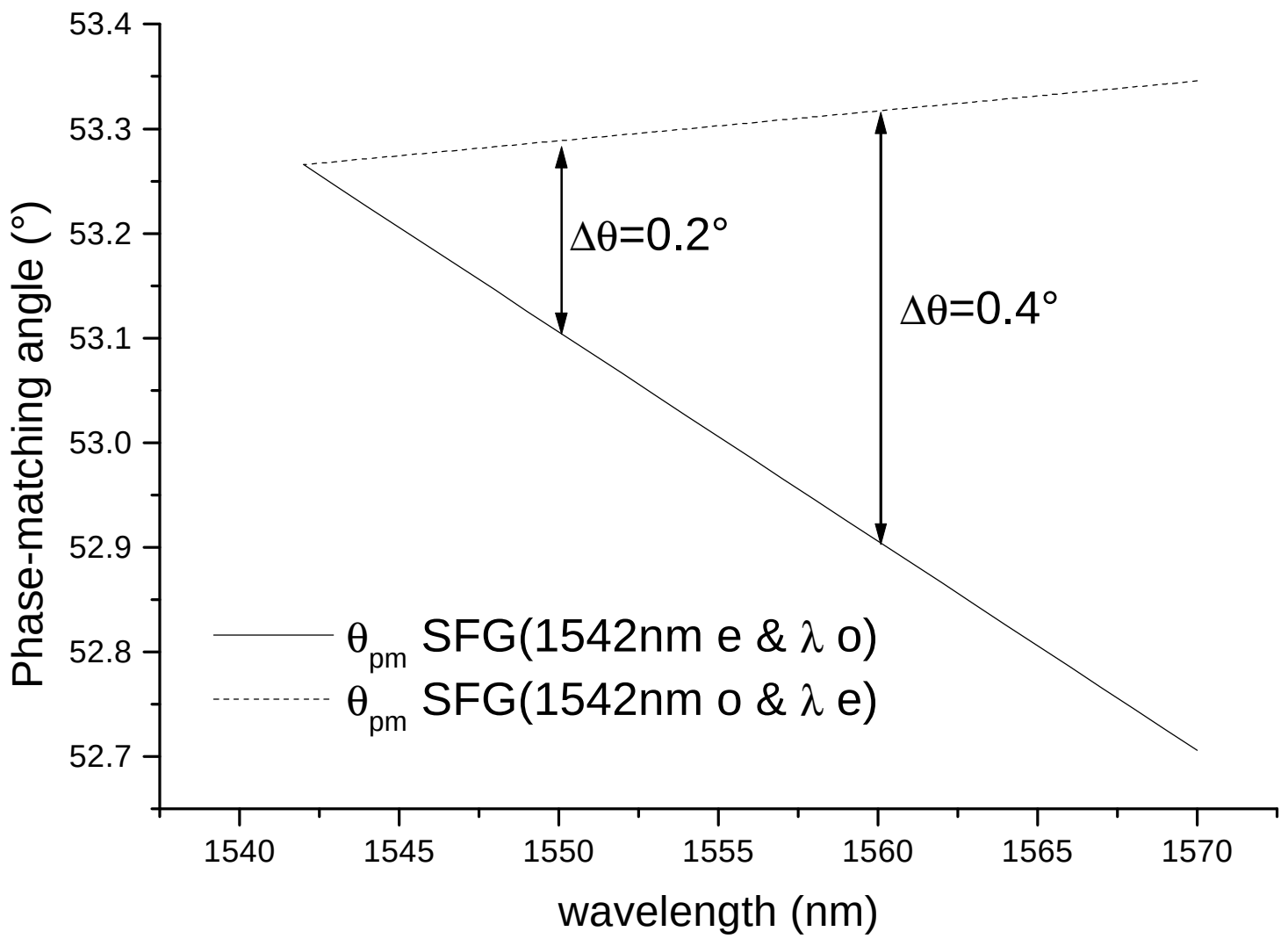

Figure 3

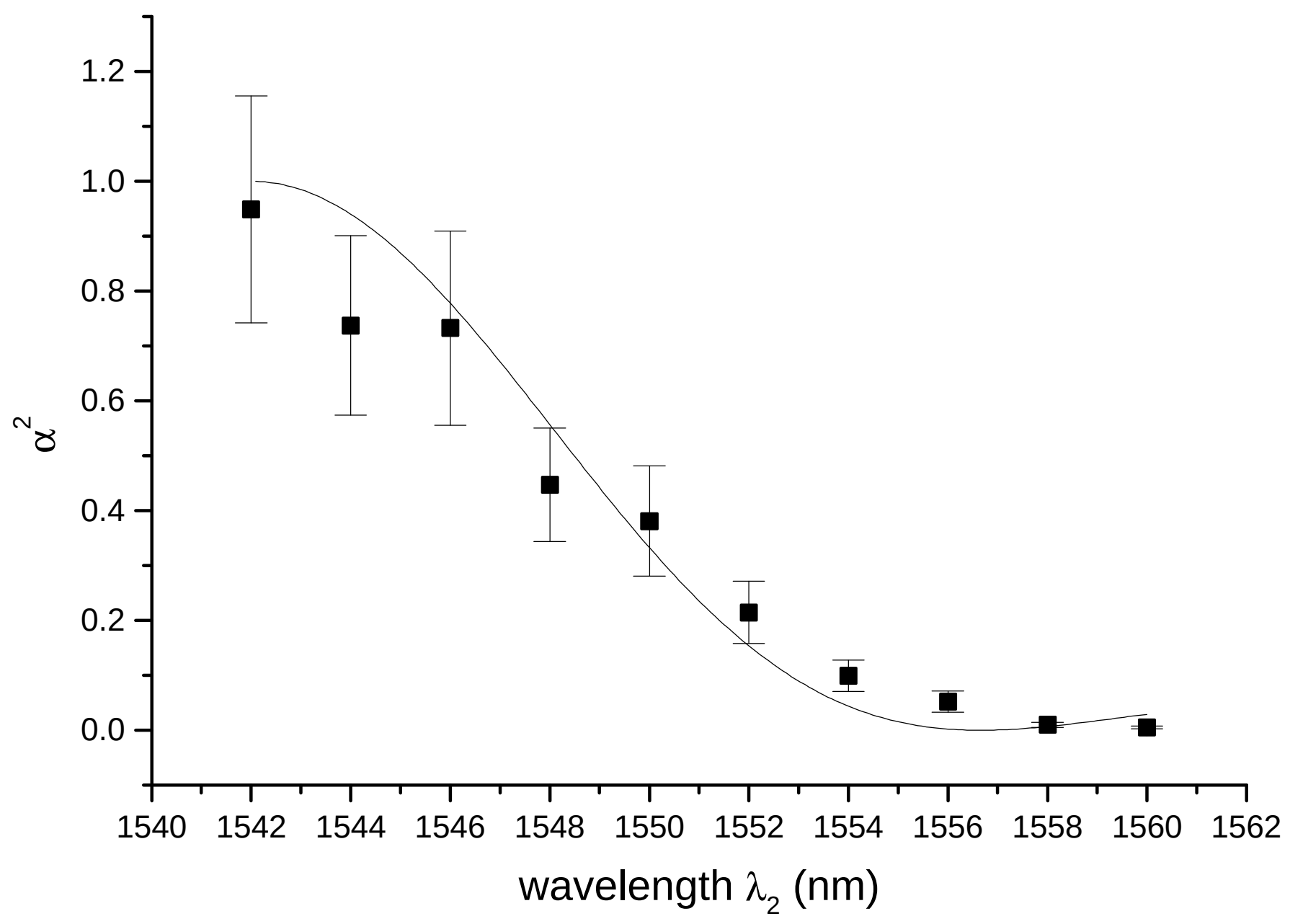

1.2
1.0
0.8
0.6
0.4
0.2
0.0
$\alpha^2$
1540 1542 1544 1546 1548 1550 1552 1554 1556 1558 1560 1562
wavelength $\lambda_2$ (nm)

Figure 4

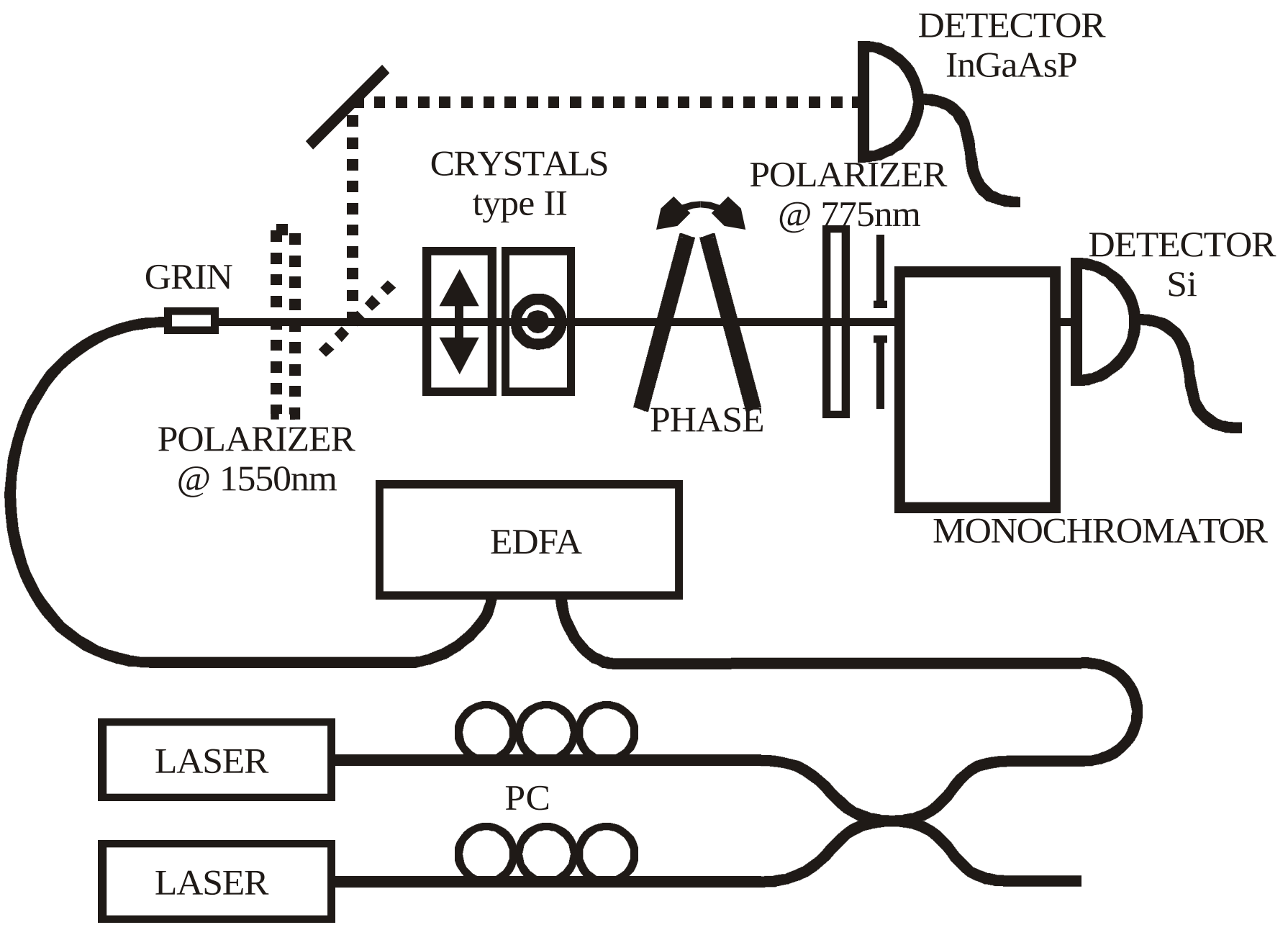

Figure 5

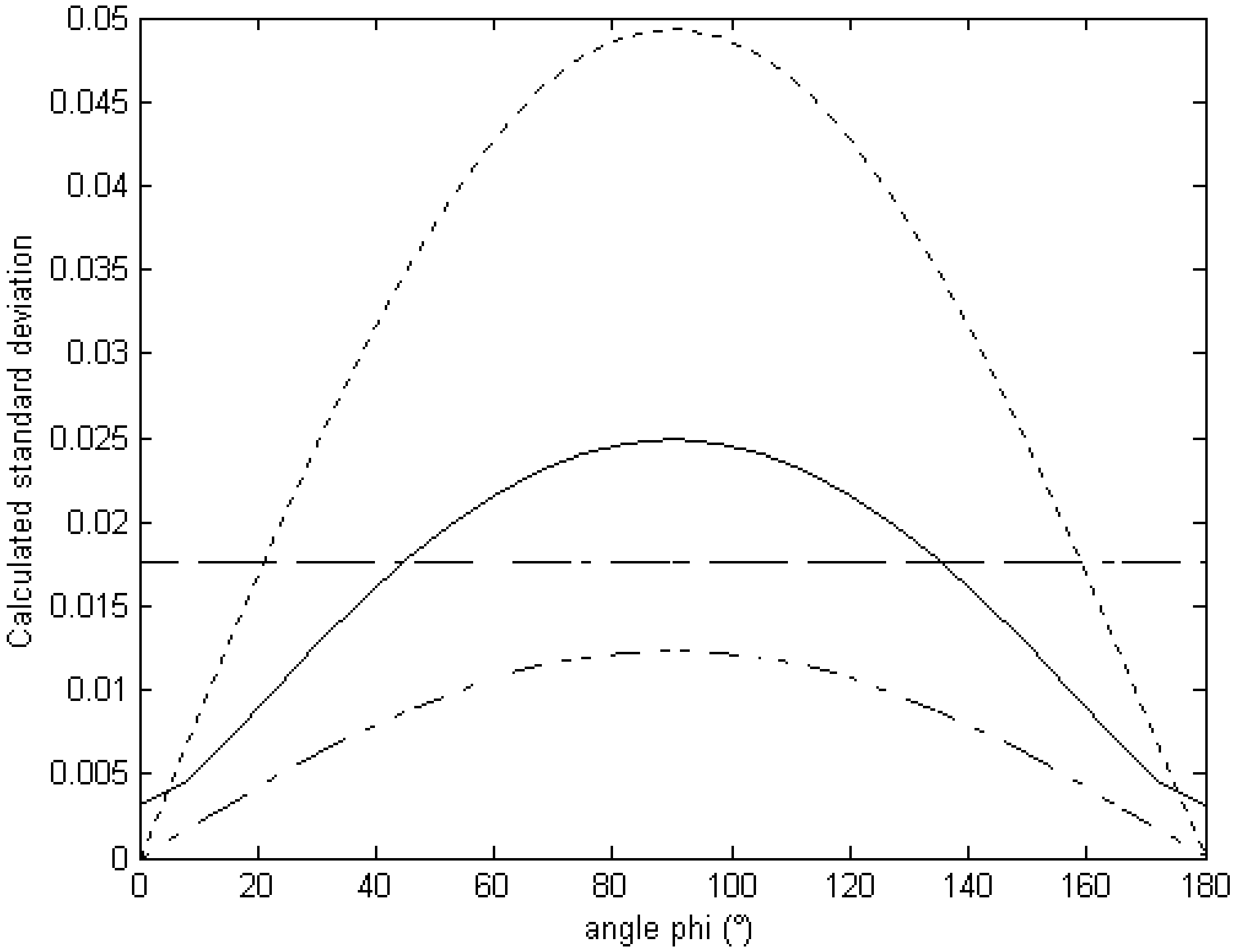

Figure 6

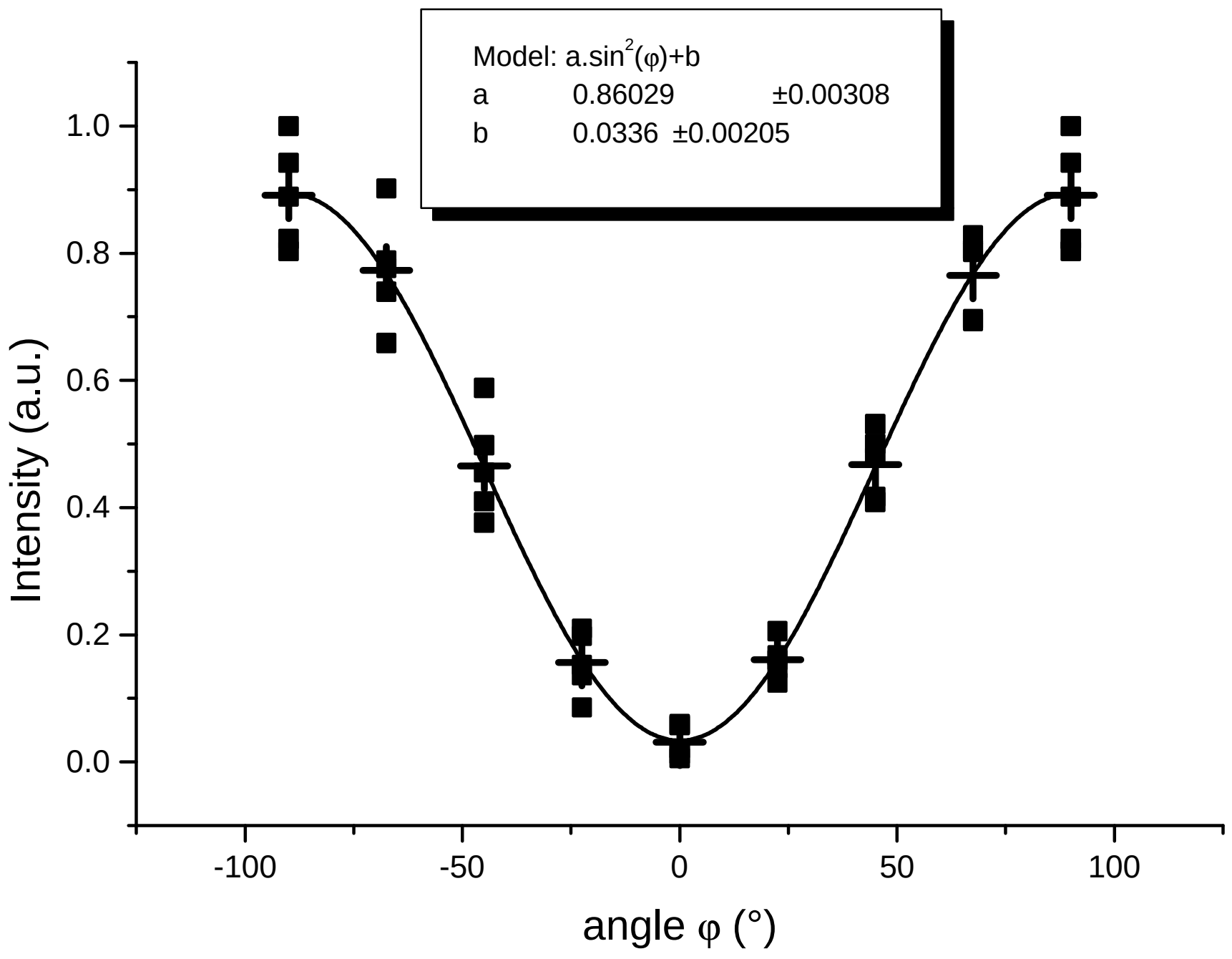

Figure 7

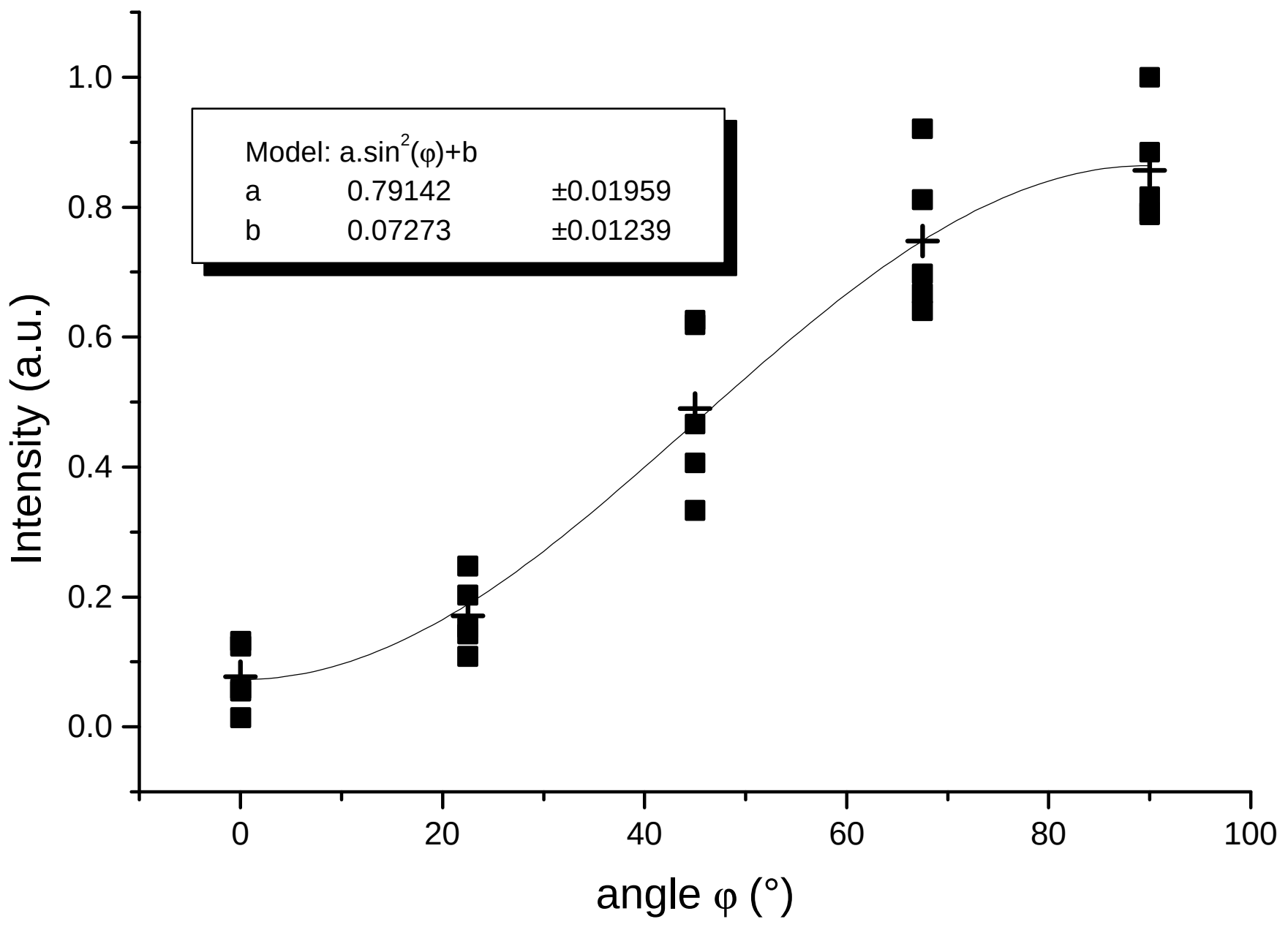

Figure 8

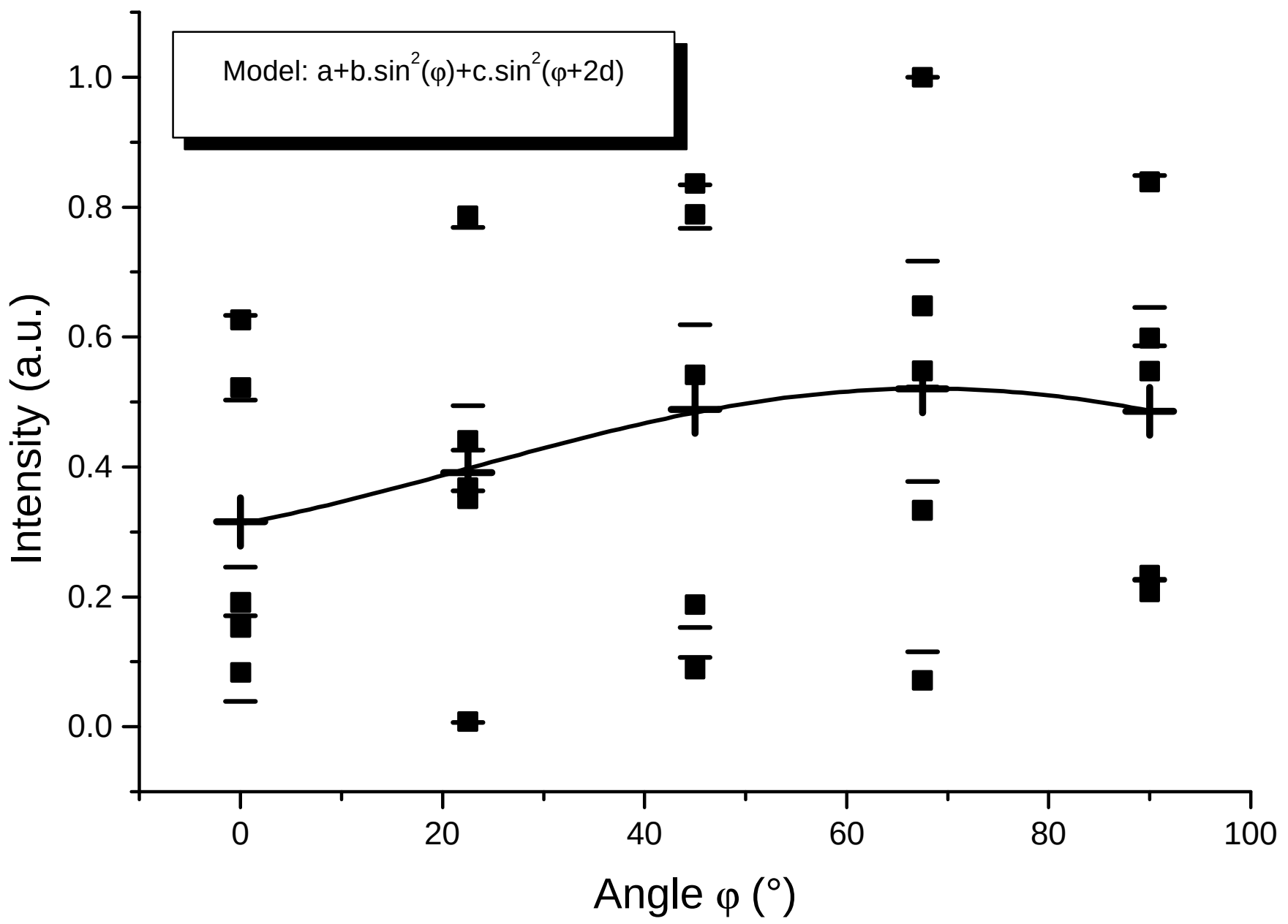

Figure 9